\documentclass[aip,reprint]{revtex4-1}
\usepackage{amsmath,graphicx}

\usepackage{color}

\newcommand{\innert}{outer }
\newcommand{\outert}{inner }
\newcommand{\cv}{l}

\begin{document}

\title{Microwave photon-assisted phase-incoherent Cooper-pair tunneling in a Josephson STM}

\author{A. Roychowdhury}
\altaffiliation{Currently: Intel Corp., Hillsboro, Oregon}
\affiliation{Laboratory for Physical Sciences, College Park, MD 20740}
\affiliation{Center for Nanophysics and Advanced Materials and Joint Quantum Institute, Dept.\ of Physics, University of Maryland, College Park, MD 20742}
\author{M. Dreyer}
\affiliation{Laboratory for Physical Sciences, College Park, MD 20740, Dept.\ of Physics, University of Maryland, College Park, MD 20742}
\author{J. R. Anderson}
\affiliation{Center for Nanophysics and Advanced Materials and Joint Quantum Institute, Dept.\ of Physics, University of Maryland, College Park, MD 20742}
\author{C. J. Lobb}
\affiliation{Center for Nanophysics and Advanced Materials and Joint Quantum Institute, Dept.\ of Physics, University of Maryland, College Park, MD 20742}
\author{F. C. Wellstood}
\affiliation{Center for Nanophysics and Advanced Materials and Joint Quantum Institute, Dept.\ of Physics, University of Maryland, College Park, MD 20742}
\date{\today}
\begin{abstract}
We have observed photon-assisted Cooper-pair tunneling in an atomic-scale Josephson junction formed between a superconducting Nb tip and a superconducting Nb sample in a scanning tunneling microscope (STM) at 30 mK. High-resolution tunneling spectroscopy data show a zero-bias conduction peak and other sharp sub-gap peaks from coupling of the STM junction to resonances in the electromagnetic environment. The sub-gap peaks respond to incident microwave radiation by splitting into multiple peaks with the position and height depending on the frequency and amplitude of the microwaves. The inter-peak spacing shows that the charge carriers are Cooper pairs, rather than quasiparticles, and the power dependence reveals that the current originates from photon-assisted phase-incoherent tunneling of pairs, rather than the more conventional phase-coherent tunneling of pairs that yields Shapiro steps. 
\end{abstract}
\maketitle

\section{Introduction}
The use of superconducting (SC) tips, instead of normal-metal tips, in scanning tunneling microscopy (STM) allows for enhanced spectroscopic resolution due to the singularity in the density of states at the SC gap edge \cite{Rodrigo04, Guillamon08, Noat10}. In addition, the ability of a SC tip to probe the pair condensate in a SC sample on the atomic scale has inspired recent interest in Josephson STMs \cite{Rodrigo06, Kohen05, Bergeal08, jaeck01}. However, pioneering work at 2.1 K with SC tips and samples \cite{Naaman01} has revealed a resistive zero-bias conductance peak (ZBCP), rather than a true phase-coherent Josephson supercurrent \cite{josephson62}, due to classical phase diffusion that is governed by the physics of ultra-small Josephson junctions (small capacitance and small critical current)\cite{Anchenko69, Kimura08}.

In this paper, we present phase-incoherent Cooper pair tunneling data obtained at millikelvin temperatures in a superconducting Nb-Nb STM junction. Although the tunneling is phase incoherent, we show that the charge of the carriers of 2e can be unambiguously determined by applying microwaves to produce photon-assisted tunneling. Since the tunneling current arises from an atomic scale region, in principle the technique allows the discrimination of normal regions in highly inhomogeneous SC samples \cite{Wolf94,Howald01,Fischer01}, the unambiguous detection of small SC regions in otherwise normal metal samples, and the independent determination of the supercurrent fraction of a localized zero bias conductance peak or other features that occur in tunneling spectroscopy. 

\section{Theoretical considerations}
Photon-assisted \textit{quasiparticle} tunneling has been studied extensively in thin-film super\-con\-duc\-tor-insulator-superconductor (S-I-S) tunnel junctions and single electron transistors \cite{Tien63, Sweet70, Kouwenhoven94, Hergenrother94, Nakamura96, Fitzgerald98, de_Graaf13}. In a junction driven by microwaves of frequency $\omega$, the time-averaged (dc) quasiparticle current through the junction is given by \cite{Falci91}
\begin{equation}\label{eq:iqp}
\overline{I_\text{qp}(V_0,V_\mu)} = \sum_{\cv= -\infty}^\infty J_\cv^2\left(\frac{eV_{\mu}}{\hbar \omega}\right) I_\text{qp}\left(V_0 - \frac{\cv\hbar \omega}{e}\right),
\end{equation}
where $V_{\mu}$ is the amplitude of the applied microwaves seen by the junction, $V_0$ is the dc bias voltage across the junction, $J_\cv$ is the $\cv^{th}$ Bessel function, and $I_\text{qp}(V)$ is the quasiparticle current when no microwave voltage is applied. In contrast, the phase-incoherent, time-averaged Cooper pair current through an ultra-small junction that is driven by microwaves at frequency $\omega$ is given by \cite{Falci91}
\begin{equation}\label{eq:isc}
\overline{I_\text{s}(V_0,V_\mu)} = \sum_{\cv=-\infty}^\infty J_\cv^2\left(\frac{2eV_{\mu}}{\hbar \omega}\right) I_\text{s}\left(V_0 - \frac{\cv\hbar \omega}{2e}\right),
\end{equation}
where $I_\text{s}(V)$ is the phase-incoherent Cooper pair current in the absence of microwaves, and $V_\mu$, $V_0$ and $J_\cv$ have the same meaning as in the quasiparticle case.

Phase-incoherent pair tunneling requires an ultra-small SC junction, subject to fluctuations that destroy phase coherence \cite{Tinkham}. This limit is easily obtained in an STM junction because of the typically small junction capacitance $C<1$ fF and critical current $I_\mathrm{c}<1$ nA. In contrast, the much larger critical current ($\mu$A) and capacitance (pF) of typical macroscopic Josephson junctions produces phase-coherent tunneling and inhibits phase-incoherent tunneling of Cooper pairs. 

Microwaves incident on a phase coherent junction produce Shapiro steps due to synchronization of phase oscillations with the incident microwaves \cite{Shapiro63}. For a voltage-biased macroscopic junction with critical current $I_\mathrm{c}$ the time-dependent supercurrent is given by \cite{Tinkham}
\begin{multline}\label{eq:iss}
I_\text{s}(V_0,t) =\sum_{\cv=-\infty}^\infty (-1)^\cv J_\cv\left(\frac{2eV_{\mu}}{\hbar\omega}\right)I_\text{c}\\ \cdot\sin\left[\gamma_0 + \left(\frac{2eV_0}{\hbar}-\cv\omega\right)t\right].
\end{multline}
Note that when $2eV_0=\cv\hbar\omega$, the time dependence disappears, leaving dc-supercurrent Shapiro steps with amplitude $2J_\cv\left(\frac{2eV_\mu}{\hbar\omega}\right) I_\mathrm{c}$. While Eq.\ (\ref{eq:iss}) is superficially similar to Eq.\ (\ref{eq:isc}), the differences are significant [for example, $J_\cv$ \textit{vs.} $J_\cv^2$ and $I_\mathrm{c}$ \textit{vs.} $I_s\left(V_0-\cv\frac{\hbar\omega}{2e}\right)$], making it possible to experimentally distinguish phase coherent and incoherent tunneling.

\begin{figure}[t]
\centering
\includegraphics[width=0.99\columnwidth]{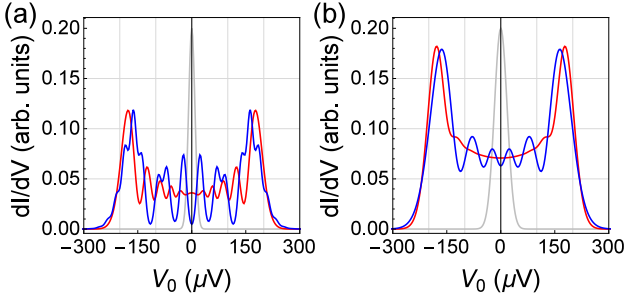}%
\caption{Simulated splitting of a single Gaussian zero-bias conductance peak (gray) due to microwaves when the charge carriers are Cooper-pairs (red) or quasiparticles (blue). The microwave frequency and amplitude were $\omega/2\pi=5.6$ GHz ($\hbar\omega/e\approx 23$ $\mu$eV) and $V_\mu=200$ $\mu$V for both plots. The width of the peak was chosen as 12.5 $\mu$V (a) and 25 $\mu$V (b), respectively. The height of each ZBCP is scaled by a factor of $1/5$ to fit in the same plot. \label{fig:sim}}
\end{figure}
\begin{figure}[t]
\centering
\includegraphics[width=0.9\columnwidth]{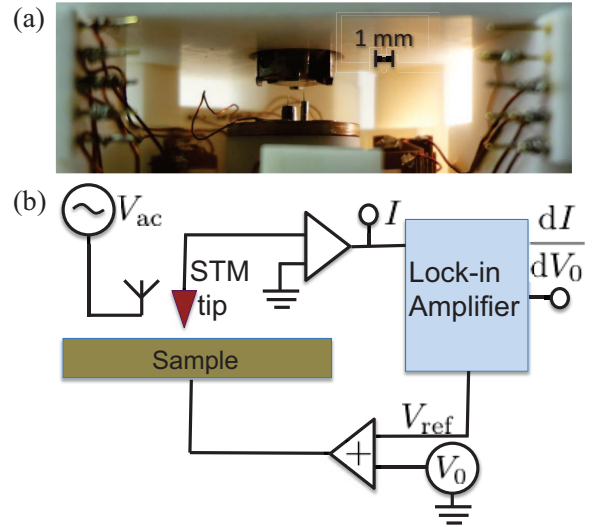}
\caption{(a) Photograph of the STM with two Nb tips and a Nb(100) sample. (b) Simplified schematic of experimental set up \cite{anita-thesis}. $V_\text{ac}$ is the amplitude of the applied microwaves at the source, I is the tunnel current output, $V_0$ is the dc bias voltage, and $V_\text{ref}$ is a 1.973 kHz sinusoidal reference from the lock-in amplifier. The coupling of the microwaves to the STM tip is represented as an antenna.\label{fig:exp}}
\end{figure}
\begin{figure}[t]
\centering
\includegraphics[width=0.9\columnwidth]{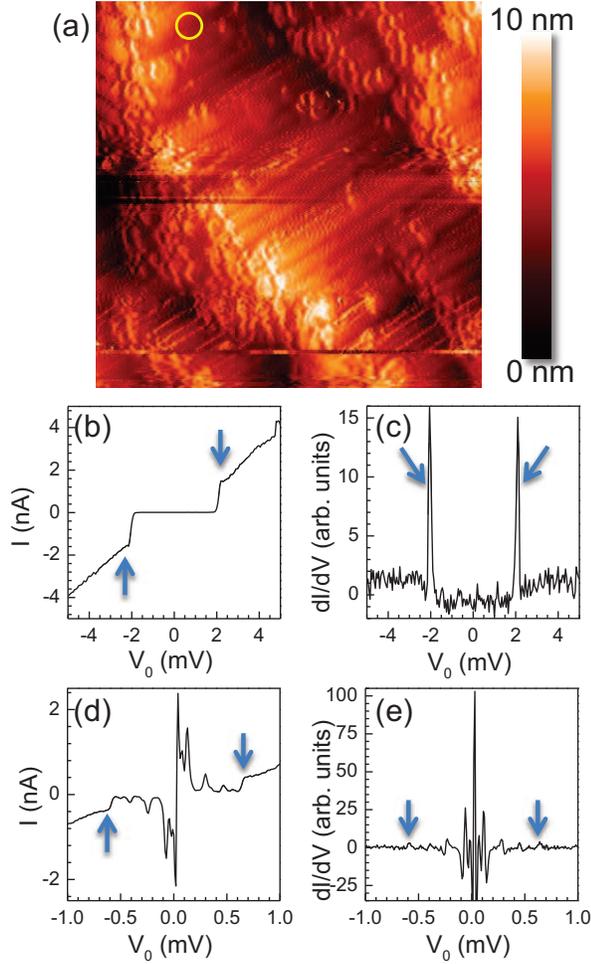}
\caption{(a) Typical topography of the Nb(100) surface using a Nb tip at 30 mK. The image size is $25\times 25$ nm$^2$ with a corrugation of 6 nm and an rms roughness of 0.98 nm. (b) I(V) and (c) dI/dV data, respectively, showing the full SIS gap (arrows, $\Delta_\mathrm{SIS}=2.08$ meV) at a junction resistance of $R_J=16.7$ M$\Omega$. Curves (d) and (e) show I(V) and dI/dV data, respectively, of the fine sub-gap structure. Here the arrows mark the position of the smaller of the two gaps, most likely the tip gap, of $\Delta_\mathrm{Tip}=0.61$ meV at $R_J=10.0$ M$\Omega$ implying $\Delta_\mathrm{Sample}=1.47$ meV. The yellow circle in (a) marks the region where the spectroscopic data were acquired.\label{fig:toposis}}
\end{figure}
\begin{figure}[h]
\centering
\includegraphics[width=0.75\linewidth]{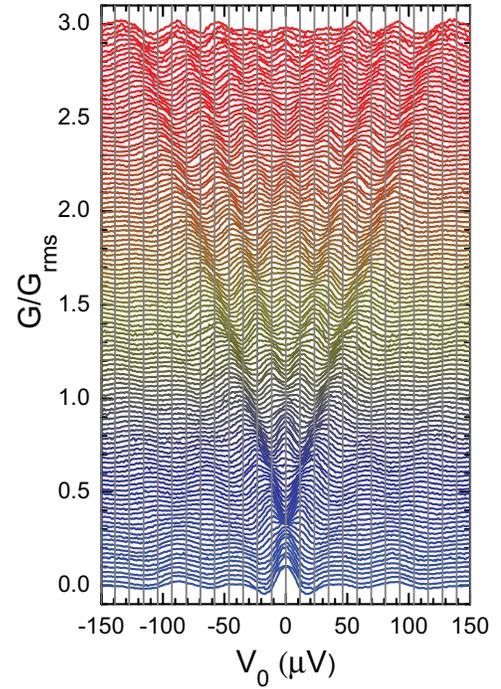}
\caption{Normalized measured conductance $G/G_\text{rms}$ versus dc bias voltage $V_0$ taken at 30 mK with the \outert Nb-Nb STM junction under irradiation by 5.6 GHz microwaves. The amplitude of the applied voltage $V_\text{ac}$ varies from $0~V$ (blue) to $3.0~V$ (red) in steps of 25 mV. Successive curves are offset by 0.025 on the $y$-axis. Vertical gray lines are spaced $\hbar \omega / 2e = 11.6\ \mu$V apart and coincide with emerging peaks in conductance, indicating that the charge of the carriers is $2e$. $G_\text{rms}$ is the rms deviation of $G$ of each trace.\label{fig:plot56}}
\end{figure}
\begin {figure}[t]
\includegraphics[width=\linewidth]{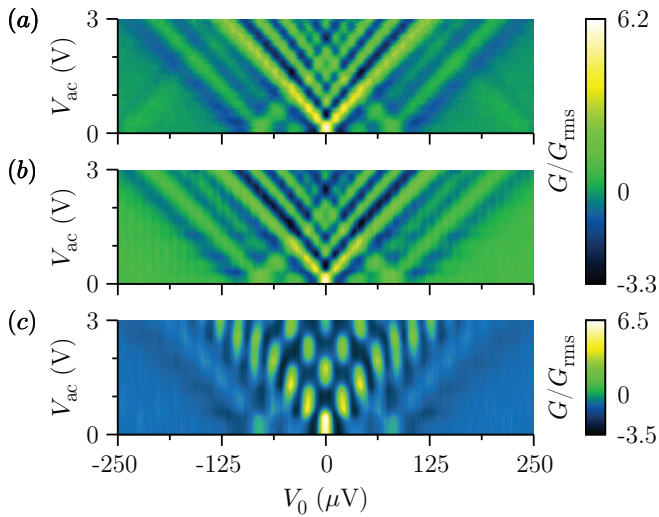}
\caption{(a) False color plot of data in Fig.~\ref{fig:plot56} showing measured conductance $G/G_\text{rms}$ versus dc bias voltage $V_0$ and applied microwave amplitude $V_\text{ac}$ at a frequency of $f=5.6$ GHz. (b) Simulated false color plot assuming charge carriers are Cooper pairs. The measured conductance curve in the absence of microwaves and Eq.\ (\ref{eq:gsc}) were used to generate each successive curve, with $A_{\mu}=V_{\mu}/V_\text{ac} \approx 6.5 \times 10^{-5}$. (c) Simulated false color plot assuming the charge carriers are quasiparticles with charge $e$. The measured conductance curve in the absence of microwaves and the equivalent of Eq.\ (\ref{eq:iqp}) were used to generate each successive curve with $A_{\mu} = 6.5\times 10^{-5}$.\label{fig:img56}}
\end{figure}
\begin {figure}[t]
\includegraphics[width=0.75\linewidth]{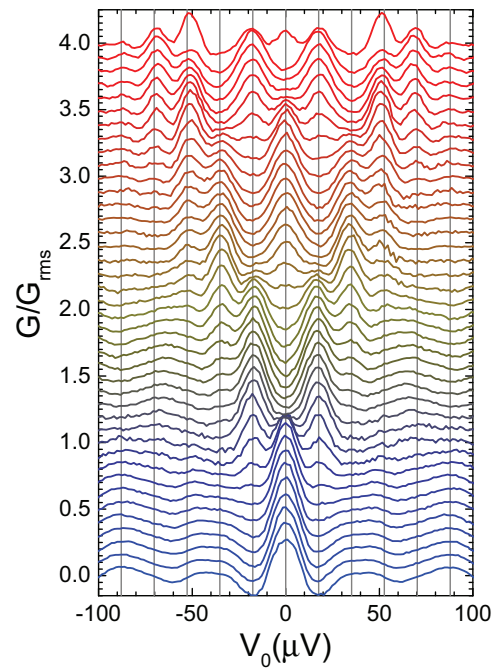}
\caption{Measured conductance $G/G_\text{rms}$ versus dc bias voltage $V_0$ taken at 30 mK with the \outert tip Nb-Nb STM junction under irradiation of $f=8.5$ GHz microwaves. The amplitude of the applied microwaves $V_\text{ac}$ is varied from $0~V$ (blue) to $4.0~V$ (red) in steps of 0.1 V. Successive curves are offset by 0.1 on the y-axis. Vertical gray lines are spaced $\hbar\omega/2e = 17$ $\mu$eV apart. They coincide with high peaks in conductance, indicating that the charge of the carriers is $2e$.\label{fig:plot85}}
\end{figure}
\begin {figure}[t]
\centering
\includegraphics[width=\linewidth]{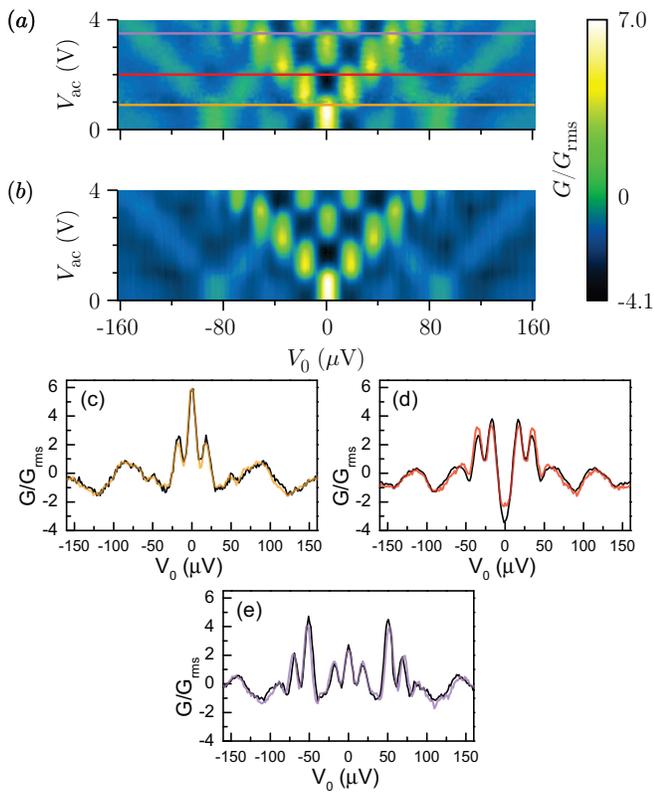}
\caption{(a) False color plot of Fig.~\ref{fig:plot85} showing measured conductance $G/G_\text{rms}$ versus dc bias voltage $V_0$ and applied microwave amplitude $V_\text{ac}$ for a microwave frequency of $f=8.5$~GHz. (b) Simulated false color plot for phase-incoherent pair tunneling generated using the measured conductance curve in the absence of microwaves and Eq.\ (\ref{eq:gsc}) to generate each successive curve with $A_\mu = V_\mu/V_\text{ac} \approx 3.5 \times10^{-5}$. (c) Orange curve shows $G/G_\mathrm{rms}$ at $V_\text{ac} = 0.9$ V, and black shows corresponding simulated curve from (b). (d)~Red curve shows data at $V_\mathrm{ac} = 2.0$ V, black shows corresponding simulated curve from (b). (e) Purple curve shows data at $V_\mathrm{ac} = 3.5$ V, and black curve is the corresponding simulated curve from (b).\label{fig:cmp85}}
\end{figure}

In practice we measure the dc conductance $G(V_0,V_\mu)=\mathrm{d}\overline{I}/\mathrm{d}V_0$ and obtain from Eq.~(\ref{eq:isc}) for example
\begin{subequations}
\begin{equation}\label{eq:gsc}
G_\text{s}(V_0,V_\mu) = \sum_{\cv=-\infty}^\infty J_\cv^2\left(\frac{2 e V_{\mu}}{\hbar \omega}\right) G_\text{s}\left(V_0 - \frac{\cv\hbar \omega}{2 e},0\right)
\end{equation}
When Eq.~(\ref{eq:iqp}) is used as a starting point we obtain
\begin{equation}\label{eq:gqp}
G_\text{qp}(V_0,V_\mu) = \sum_{\cv=-\infty}^\infty J_\cv^2\left(\frac{e V_{\mu}}{\hbar \omega}\right) G_\text{qp}\left(V_0 - \frac{\cv\hbar \omega}{e},0\right)
\end{equation}
\end{subequations}
In Eqs.~(\ref{eq:gsc}) and (\ref{eq:gqp}), the sum over $\cv$ may be interpreted as microwaves causing an energy level $\epsilon$ to split into levels $\epsilon\pm \cv\hbar\omega$, corresponding to the absorption or emission of $\cv$ photons of frequency $\omega$. The equations also imply that a slowly-varying $G(V_0,0)$ \textit{vs.} $V_0$ curve will exhibit little change in appearance due to microwave radiation. On the other hand, a $G(V_0,0)$ \textit{vs.} $V_0$ characteristic with features that are sharp compared to $\hbar \omega /ne$ will respond under microwave radiation by developing features shifted by $\hbar \omega/ne$ along the voltage axis, where $ne$ is the charge of the carriers. Thus, for tunneling Cooper pairs, a sharp feature in the $G(V_0,0)$ curve will be shifted in voltage by increments of $\hbar\omega /2e$ when microwaves are applied, or by twice this spacing in the quasiparticle case.

It should be pointed out that the range $\pm V_0$ for which the junction responds to microwaves is not a function of carrier charge or frequency but rather is approximately equal to the microwave amplitude. It is therefore necessary to resolve the fine structure to distinguish the charge. Figure \ref{fig:sim}(a) shows a simulation of splitting a sharp ZBCP. Here the 12.5 $\mu$V peak width is half of the microwave induced energy level spacing for quasiparticles. The difference between pair tunneling (red) and quasiparticle tunneling (blue) is clearly visible. Figure \ref{fig:sim}(b) shows a borderline case where the width of the ZBCP is comparable to the quasiparticle levels. In (b) a presumed Cooper-pair current would no longer show a clear split-peak structure while the quasiparticle current still does. A slight increase in peak width would wipe out this structure making the cases virtually indistinguishable.  A sharp peak and a high energy resolution compared to the microwave frequency are thus necessary to distinguish the carrier charge. In addition, by fitting to the weighted sum of the supercurrent and quasiparticle current, i.e.:
\begin{equation}\label{eq:gscqp}
G(V_0,V_\mu) = a_\text{qp} G_\text{qp}(V_0,V_\mu)+a_\text{sc} G_\text{sc}(V_0,V_\mu)
\end{equation}
it is possible to determine the quasiparticle and supercurrent fractions when both carriers are present. In simulations we have found that fractions as low as 0.01 \% can easily be discerned. In practice, the noise level of the data determines the detection limit. For all the data shown in this paper the quasiparticle current fraction is below the detection limit of $\approx 2\%$.

Fortuitously, sharp features are expected in ultra-small S-I-S junctions if the junction is connected to bias leads that have transmission line resonances or other microwave resonances. When resonances exist, theory predicts the probability $P(E)$ for energy $E$ to be transferred from the tunneling charges to the circuit \cite{Averin86, Ingold94}, leading to conductance peaks. To achieve the energy resolution necessary to observe these fine-scaled features and the response to microwaves, we cool the STM \cite{Roychowdhury14} to 30 mK.

\section{Experimental Setup}
Figure \ref{fig:exp} shows a photograph of our STM and a simplified schematic of our measurement set up. The STM has two independent tips (``inner'' and ``outer'') and is mounted on a custom Oxford Instruments dilution refrigerator \cite{Roychowdhury14}. Both tips were cleaned by high voltage field emission on a gold single crystal at low temperatures before changing to the Nb sample. Microwave power \footnote{Agilent N5183A} was transmitted indirectly to the STM tips \textit{via} a dc thermometer line. A lock-in amplifier \footnote{Stanford Research Systems SR830 lock-in amplifier} was used to measure the conductance ($G=\mathrm{d}\overline{I}/\mathrm{d}V_0$) as a function of the dc bias $V_0$. The bulk Nb (100) sample was prepared by heating it to $600^\circ$ C in ultra-high vacuum for 10-12 hours at a time, while sputtering it with 2 keV Ar$^+$ ions for 9 consecutive days. Once residual polishing grains had been removed, the sample was sputtered with 1 keV Ar$^+$ ions at a temperature of 600 $^\circ$C for 2-3 hours before transferring it to the STM without breaking vacuum \cite{Roychowdhury14}.

A topographic image of the resulting surface is shown in Fig.~\ref{fig:toposis}(a). Since the maximum heating temperature was relatively low, the sample does not show a clear mono-atomic step structure as one would expect from a single crystal. However, we found small flat areas (yellow circle in Fig.~\ref{fig:toposis}(a)) to conduct spectroscopic measurements. Figures \ref{fig:toposis}(b) and (c) show the SIS gap in $I(V)$ as well as $\mathrm{d}I/\mathrm{d}V$ spectroscopy. The sub-gap structure is only visible at smaller tip-sample separation as shown in Figures \ref{fig:toposis}(d) and (e). Fig.~\ref{fig:toposis}(b) and (c) allowed measurement of $\Delta_\mathrm{SIS}=\Delta_\mathrm{Tip}+\Delta_\mathrm{Sample}$, while 3(c) and (d) give the smaller of the tip and sample superconducting gap. Typically \cite{roychowdhury14-1} we expect the tip to have the smaller gap. Hence, from $\Delta_\mathrm{SIS}=2.08$ meV and $\Delta_\mathrm{Tip}=0.61$ meV, we find $\Delta_\mathrm{Sample}=1.47$ meV. Similar characterizations were performed prior to each microwave power dependent series presented in this paper.   
 
\section{Measurement and Discussion}
Figure \ref{fig:plot56} shows a series of conductance $G(V_0)$ curves taken with the \outert STM tip at the position marked in Fig.~\ref{fig:toposis}(a) with applied microwaves of frequency $\omega/2\pi=f=5.6$ GHz. Starting with zero microwave power, each successive curve was measured at a fixed microwave source amplitude ($V_\text{ac}$) that was increased in steps of 25 mV from 0 to 3 V. The bottom curve ($V_\mathrm{ac}=0$ V) shows a distinct conduction peak at zero bias, as expected, due to phase diffusion \cite{Tinkham}. The weaker side structures are due to coupling to microwave modes in the environment and can -- in principle -- be described by $P(E)$ theory \cite{Ingold94}. As the microwave amplitude increased, additional peaks appeared in the conductance curve. The position of those peaks coincide with the vertical gray lines spaced $\hbar\omega/2e = 11.6~\mu$eV apart (see Fig.\ \ref{fig:plot56}). For each conductance curve of N points, we normalized by the standard deviation $G_\text{rms}=\sqrt{1/(N-1)\sum_{n=1}^N (G_n-\overline{G})^2}$ of that curve to compensate for variations between curves due to the systematic decline in feature size for higher microwave amplitudes. 

\begin{figure}[t]
\centering
\includegraphics[width=\linewidth]{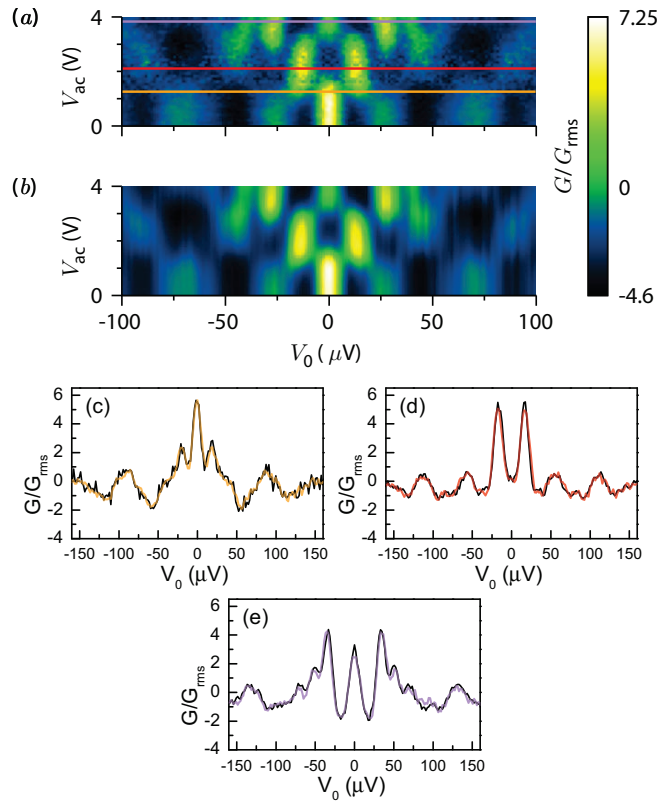}
\caption{(a) False color plot of measured conductance $G/G_\text{rms}$ versus dc bias voltage $V_0$ and applied microwave amplitude $V_\text{ac}$. Data were taken at 30 mK with \innert STM tip. The microwave frequency was $f=8.5$ GHz. Each horizontal line corresponds to a conductance curve with yellow representing the positive peaks and dark blue representing the negative dips. (b) Simulation of false color plot for phase-incoherent pair tunneling generated using the measured conductance curve in the absence of microwaves and Eq.\ (\ref{eq:gsc}) to generate each successive curve with $A_\mu = V_\mu/V_\text{ac} \approx 2.9 \times 10^{-5}$. (c)-(e) Line cuts similar to Fig.~\ref{fig:cmp85} showing data (color) and simulation (black) at $V_\mathrm{ac}=1.3$ V, 2.1 V, and 3.9 V, respectively.\label{fig:ot}}
\end{figure}

Figure \ref{fig:img56}(a) shows a false-color plot of the data displayed in Fig.~\ref{fig:plot56}, while Fig.~\ref{fig:img56}(b) shows for comparison the expected response for phase-incoherent pair tunneling based on Eq.~(\ref{eq:gsc}). To generate Fig.~\ref{fig:img56}(b), we used the first line of conductance data, measured in the absence of microwaves, and generated each successive line of non-zero microwave amplitude by applying Eq.~(\ref{eq:gsc}). Each simulated curve was divided by its standard deviation $G_\text{rms}$ to allow direct comparison with the data in Fig.~\ref{fig:img56}(a). We note that the only fitting parameter was an overall scale factor $A_\mu\equiv V_\mu/V_\text{ac}$: the ratio of the amplitude of the applied voltage across the junction $V_\mu$ to the amplitude $V_\text{ac}$ at the source. For Fig.~\ref{fig:img56}(b), we set $A_\mu = 6.5 \times 10^{-5}$. 

We see excellent agreement between Fig.\ \ref{fig:img56}(b) and Fig.\ \ref{fig:img56}(a), indicating that the current is due to phase incoherent tunneling of pairs. In particular, the charge carriers are Cooper pairs because the voltage spacing between the split conductance peaks is $\hbar \omega /2e$. The $P(E)$ structures at $V_0\approx\pm 30$ $\mu$V and $V_0\approx\pm 80$ $\mu$V split in similar fashion and are thus also due to Cooper pairs. In contrast, Fig.~\ref{fig:img56}(c) shows the corresponding simulation assuming the charge carriers are quasiparticles with charge $e$. The voltage spacing in Fig.~\ref{fig:img56}(c) is twice that for Cooper pairs, and disagrees strongly with the data.

We also measured the \outert STM junction's response to $f=8.5$ GHz microwaves at a different location on the sample ($\Delta_\mathrm{Tip}=0.625$ meV, $\Delta_\mathrm{Sample}=1.51$ meV). Since the spacing between the peaks should scale with frequency, they should be easier to resolve provided sufficient power reaches the junction. Figure~\ref{fig:plot85} shows a series of normalized conductance curves measured as the applied microwave amplitude was increased from 0 V to 4.0 V. The gray lines spaced by $\hbar \omega / 2e = 17~\mu$eV once again coincide with the measured peaks. The corresponding false color map is shown in Fig.\ \ref{fig:cmp85}(a), and the simulated false color map based on the curve measured at zero microwave power and Eq.\ (\ref{eq:gsc}) for pair tunneling is shown in Fig.\ \ref{fig:cmp85}(b). Figures \ref{fig:cmp85}(c), (d) and (e) show line sections through the data  marked by the orange, red and purple lines in Fig.~\ref{fig:cmp85}(a), and the corresponding simulated curves from Fig.~\ref{fig:cmp85}(b). The quantitative agreement is very good, consistent with the peaks being due to phase incoherent tunneling of pairs and inconsistent with quasiparticles or Shapiro steps that would arise from phase coherent tunneling of pairs.

To rule out the possibility of this being a junction-specific phenomenon, the \innert Nb tip in our dual-tip STM was used to confirm the results. Here the tip and sample gap were $\Delta_\mathrm{Tip}=1.35$ meV and $\Delta_\mathrm{Sample}=1.37$ meV, respectively. Figure\ \ref{fig:ot}(a) shows conductance measurements with the \innert STM tip with $f=8.5$ GHz microwave radiation. Close comparison of Figs.~\ref{fig:cmp85}(a) (\outert tip) and \ref{fig:ot}(a) (\innert tip) reveal small differences. Since each tip of our STM had its own set of current and piezo leads, the resonant microwave frequencies associated with each circuit are different, leading to small differences in the $V_\mathrm{ac}=0$ V conductance curve. Nevertheless, we again find very good agreement between the data and Eq.\ (\ref{eq:gsc}) [see Fig.\ \ref{fig:ot} (a)-(e)].
  
\section{Conclusion}

Sub-gap conductance features occur in voltage-biased SC STM junctions due to resonances in the junctions' electromagnetic environment. When microwave radiation is applied, the features evolve as the microwave voltage is increased. In our ultra-low temperature system, these features are due to \textit{phase incoherent} tunneling of Cooper pairs; phase coherent tunneling of Cooper pairs is not consistent with the data. Theoretical fits to the highly-resolved tunneling spectra allow us to exclude quasiparticle contributions to the tunneling current. In principle a quasiparticle fraction of 2\% or larger could be detected.

Photon assisted tunneling with a JSTM allows the atomic scale \cite{roychowdhury14-1,anita-thesis} identification of the charge of carriers that produce any sharp voltage dependent features in conductance data. This technique could be implemented in other mK-STM systems with the `simple' addition of a microwave drive and the use of superconducting tips. While traditional STMs that rely on quasiparticle tunneling provide excellent spatial maps of various materials, they are insensitive to the origin of gap states. Our Josephson STM provides similar spatial maps of materials but additionally discerns the superconducting from quasiparticle currents. 

There are several potential applications of a Josephson STM in a microwave field. It could aid in the discovery of new superconductors, as well as improve understanding of the behavior of superconductors near atomic scale perturbations. Vortex cores, small normal regions, or the effects of single magnetic spins could be probed by mapping out the quasiparticle and Cooper pairs contributions to the current at the boundary of normal and SC regions of the samples. In addition, pseudogap states or other competing orders can be distinguished from superconductivity in spatially inhomogeneous highly correlated electron systems. Furthermore, this technique could be used to discern whether the zero-bias conductance peak in a topological superconductor arises from superconductivity or something more exotic such as the Majorana fermion\cite{RevModPhys.87.137}. Finally, the measurement technique may also provide a way to attain position-dependent measurements of local resonant absorption peaks, of interest when studying the effects of adsorbed molecules or resonant two-level systems \cite{osborn1,osborn2,osborn3} in quantum computing applications.

\begin{acknowledgments}
The authors would like to acknowledge many useful conversations with P.\ Barbara, B.\ Palmer and B.\ Suri. Portions of this work were funded by NSF under DMR-0605763.  
\end{acknowledgments}

\bibliography{pat_bibliography}

\end{document}